# Observation of nonlinear planar Hall effect in magnetic insulator/topological insulator heterostructures

Yang Wang[1], Sivakumar V. Mambakkam[2], Yue-Xin Huang[3], Yong Wang[2], Yi Ji[1], Cong Xiao[4,5], Shengyuan A. Yang[3], Stephanie A. Law[1,2], and John Q. Xiao[1,*]

[1] *Department of Physics and Astronomy, University of Delaware, Newark, Delaware 19716, USA*
[2] *Department of Materials Science and Engineering, University of Delaware, Newark, Delaware, 19716, USA*
[3] *Research Laboratory for Quantum Materials, Singapore University of Technology and Design, Singapore 487372, Singapore*
[4] *Department of Physics, The University of Hong Kong, Hong Kong, China*
[5] *HKU-UCAS Joint Institute of Theoretical and Computational Physics, Hong Kong, China*

[*]jqx@udel.edu

**Abstract**

**Interfacing topological insulators (TIs) with magnetic insulators (MIs) has been widely used to study the interaction between topological surface states and magnetism. Previous transport studies typically interpret the suppression of weak antilocalization or appearance of the anomalous Hall effect as signatures of magnetic proximity effect (MPE) imposed to TIs. Here, we report the observation of nonlinear planar Hall effect (NPHE) in $Bi_2Se_3$ films grown on MI thulium and yttrium iron garnet (TmIG and YIG) substrates, which is an order of magnitude larger than that in $Bi_2Se_3$ grown on nonmagnetic gadolinium gallium garnet (GGG) substrate. The nonlinear Hall resistance in TmIG/$Bi_2Se_3$ depends linearly on the external magnetic field, while that in YIG/$Bi_2Se_3$ exhibits an extra hysteresis loop around zero field. The magnitude of the NPHE is found to scale inversely with carrier density. We speculate the observed NPHE is related to the MPE-induced exchange gap opening and out-of-plane spin textures in the TI surface states, which may be used as an alternative transport signature of the MPE in MI/TI heterostructures.**

## I. INTRODUCTION

Magnetic topological insulators [1] have been intensively studied in the past decade, because scientifically fundamental as well as technologically promising phenomena like quantum anomalous Hall effect (QAHE) [2] and topological magnetoelectric effect [3] can arise in the surface states of magnetic TIs. So far doping magnetic elements like Cr or V [4,5] into the TI $(Bi,Sb)_2Te_3$ has been the most developed and robust method to observe QAHE [6,7]. Another approach to magnetize the topological surface states (TSS) is to proximately couple the TI to a magnetic material [8-11], preferentially an insulator, so there is no doping or current shunting effect. Although robust anomalous Hall effect (AHE) has been reported in $(Bi,Sb)_2Te_3$ coupled to MIs like TmIG [12] or $Cr_2Ge_2Te_6$ [13], weak or absence of AHE is more commonly seen in $Bi_2Se_3$ or $Bi_2Te_3$ when grown or transferred on top of MI substrates, and the suppression of weak antilocalization (WAL) is often taken as a qualitative signature of the MPE [14-17]. Conventional surface-sensitive spectroscopic methods like angle-resolved photoemission spectroscopy (ARPES) cannot be applied to probe the band structure at the buried MI/TI interface, and x-ray- or neutron-based measurements have not been successful in detecting any induced magnetic moments in TIs grown on MIs [18-20]. Therefore, searching an alternative and more sensitive transport probe of the MPE will be helpful to clarify and understand the interaction between magnetism and TSS in MI/TI heterostructures.

In recent years, planar Hall effect (PHE) in both linear [21] and nonlinear [22] regimes have been reported in nonmagnetic TIs. The linear PHE was interpreted as a result of magnetic-field-induced anisotropic backscattering [21] or magnetic-field-induced tilting of the Dirac cone with particle-hole (p-h) asymmetry [23,24]. The nonlinear PHE was attributed to the distortion or tilting of the Dirac dispersion with higher-order $k$ terms like p-h asymmetry ($k^2$) or hexagonal warping ($k^3$) by the external magnetic field [22,24].

In this work, we report the observation of NPHE in $Bi_2Se_3$ (BS) films grown on magnetic $Tm_3Fe_5O_{12}$ (TmIG) and $Y_3Fe_5O_{12}$ (YIG) substrates, which is an order of magnitude larger than that in $Bi_2Se_3$ film grown on nonmagnetic $Gd_3Ga_5O_{12}$ (GGG) substrate. While the NPHE in TmIG/BS with out-of-plane (OP) easy axis shows a linear dependence on the in-plane (IP) magnetic field, the NPHE in YIG/BS with IP easy axis takes an extra hysteretic jump around zero field, reflecting the reversal of the IP YIG magnetization. The same temperature dependence of the linear-in-$B$ and hysteretic components suggests they share the same origin. The carrier density dependence of the NPHE excludes hexagonal warping as the dominant mechanism. The enhancement of the NPHE in $Bi_2Se_3$ grown on magnetic substrates as well as the sharp increase of it below 30 K indicate MPE plays a critical role. Our results suggest the NPHE may work as a convenient and sensitive transport probe of the MPE in MI/TI heterostructures.

## II. METHODS

The 8 quintuple-layer (QL) $Bi_2Se_3$ films used in this study were grown on $Tm_3Fe_5O_{12}$, $Y_3Fe_5O_{12}$, and $Gd_3Ga_5O_{12}$ substrates with very close lattice constants and smooth surfaces in a molecular beam epitaxy (MBE) system with a base pressure of $1\times10^{-9}$ Torr, following the two-step Se-buffer layer method reported in Ref. [25]. The TmIG (30 nm) and two YIG (2.5 $\mu$m and 100 nm) films with OP and IP magnetic anisotropy respectively were deposited on GGG(111) substrates by magnetron sputtering or liquid phase epitaxy method. Prior $Bi_2Se_3$ growth, the TmIG and YIG substrates were soaked in Piranha solution ($H_2SO_4:H_2O_2$=3:1) for 5 min to clean the surface [26]. After annealing the substrates at 650 °C for 30 min and cooling down to 50 °C in the growth chamber, ~2 nm thick amorphous Se and 1 nm $Bi_xSe_{1-x}$ were deposited. Then the substrate temperature was slowly ramped to 325 °C at 10 °C/min to evaporate extra Se and crystalize the first QL $Bi_2Se_3$. And the remaining 7 QL $Bi_2Se_3$ was subsequently deposited by co-evaporating Bi and Se. After cooling to room temperature, the $Bi_2Se_3$ films were capped with 5 nm $SiO_2$ layer for protection in another magnetron sputtering chamber. As shown in Fig. 1(a), streaky reflection high-energy electron diffraction (RHEED) patterns were observed from the very first QL and x-ray diffraction (XRD) results of all the $Bi_2Se_3$ films exhibit clear (0,0,3$n$) peaks, indicating the $c$-axis growth orientation and the high and similar crystalline quality.

After the growth, the Bi$_2$Se$_3$ films were fabricated into 200×100 $\mu$m Hall bar devices by standard photolithography method and contacted with Ti/Au electrodes. The magneto-transport measurements were carried out in a home-built cryogenic system with base temperature 4.5 K. Low-frequency lock-in technique was used to detect the first- and second-harmonic longitudinal and transverse voltages.

## III. RESULTS AND DISCUSSION
### A. Linear transport properties

Fig. 2(a) displays the temperature dependence of the longitudinal resistance of four Bi$_2$Se$_3$ devices made on TmIG, YIG and GGG substrates. All of them show metallic behavior above 50 K. Below 50 K, the resistance upturn in TmIG and YIG/BS samples is more pronounced compared to that of the GGG/BS sample, especially in the YIG2/BS sample with a lower carrier density. Such insulating behavior may result from the suppression of WAL due to MPE [27] and is consistent with previous reports on iron garnet/Bi$_2$Se$_3$ bilayers [14,16,17]. As shown in Fig. 2(b), from ordinary Hall effect (OHE) measurements we extracted the sheet carrier densities for these devices to be 2.45-3.46×10$^{13}$ cm$^{-2}$. This means these Bi$_2$Se$_3$ samples are $n$-doped with a Fermi level of ~0.3 eV, so significant amount of the current is carried by the bulk states. We did not observe hysteretic or nonlinear AHE in the TmIG or YIG/BS samples after subtracting the linear OHE background. This suggests that the garnet/Bi$_2$Se$_3$ interfaces formed here may not be as good as those in Ref. [17], where weak AHE was observed. This does not rule out the existence of a small exchange-interaction-induced gap in the TSS in our samples, because when the gap $\Delta$ is small, e.g. ~1 meV, and it is much smaller than the Fermi energy $\varepsilon_\text{F}$, the AH conductivity $\sigma_{yx}^\text{AH} \propto \frac{8e^2}{\hbar}\left(\frac{\Delta}{\varepsilon_\text{F}}\right)^3$ [28] will give rise to an AH resistance in the order of 0.1 m$\Omega$, which cannot be discerned from the large OHE background. Although AHE was not detected, we observed suppression of WAL in Bi$_2$Se$_3$ films grown on TmIG and YIG substrates as compared with that on GGG [Fig. 2(c)]. By fitting to the Hikami-Larkin-Nagaoka equation, the electron phase coherence of length of the three Bi$_2$Se$_3$ samples on GGG, TmIG, and YIG are 203, 119, and 90 nm, respectively. Given the similar crystalline quality [Fig. 1] and carrier densities [Fig. 2(b)], the suppressed WAL in TmIG and YIG/BS is most likely due to MPE-induced OP spin textures and correspondingly, the reduced Berry phase of TI surface electrons [30].

### B. Observation of the NPHE

The linear PHE has $\sin\phi_B \cos\phi_B$ dependence on the IP magnetic field direction [21,23] where $\phi_B$ is the angle between the current and magnetic field directions. Thus, the first-order Hall voltage is zero at $\phi_B = n90°$ with $n$ being an integer. Differently, the NPHE depends on the IP magnetic field as $B\cos\phi_B$ [22,24], so it can be detected by sweeping the magnetic field between 0 and 180°. As illustrated in Fig. 3(a) inset, in our experiments we sent a sinusoidal a.c. current to the Hall channel in the $x$-direction, and measured the second harmonic Hall voltage $V_y^{2\omega}$ while sweeping the external magnetic field also in the $x$-direction. As shown in Fig. 3, the second harmonic Hall resistance $R_{yx}^{2\omega} = V_y^{2\omega}/I$ as a function of $B$ for three Bi$_2$Se$_3$ devices fabricated on GGG, TmIG, and YIG substrates have dramatically different behaviors. Compared with the TmIG and YIG/BS samples, the magnitude of the NPHE in GGG/BS measured by the slope $dR_{yx}^{2\omega}/dB$ is an order of magnitude smaller. In the TmIG/BS bilayer with OP magnetic anisotropy, the NPHE is enhanced and exhibits a linear dependence on the IP magnetic field [Fig. 3(b)]. When $Bi_2Se_3$ is grown on YIG substrate with IP anisotropy, $R_{yx}^{2\omega}$ not only exhibits a linear dependence on $B$, but also takes a hysteretic shape around zero field, corresponding to the switching of the IP magnetization of YIG. Moreover, the NPHE in both TmIG and YIG/BS samples is only observable below ~30 K and shows sharp increase when temperature is decreased from 30 K to 4.5 K. Such enhanced NPHE with similar characteristics was observed in different YIG/$Bi_2Se_3$ samples grown at different times and in different TmIG/$Bi_2Se_3$ devices, confirming the reproducibility of the results.

The NPHE previously reported in $Al_2O_3$/$Bi_2Se_3$ was attributed to the nonlinear, i.e., the hexagonal warping $k^3$ and the p-h asymmetry $k^2$ terms in the topological surface dispersion [22]. Although this contribution should exist in all three samples shown here, it cannot account the enhancement of the NPHE in TmIG and YIG/BS. When scaled by the coefficient $\gamma_y \equiv \frac{R_{yx}^{2\omega}}{R_{xx}IB}$, the magnitude of the NPHE in TmIG and YIG/BS at 4.5 K is 0.025 and 0.031 respectively, which is one and two orders of magnitude larger than that in GGG/BS ($\gamma_y = 3\times10^{-3}$) and $Al_2O_3$/BS ($\gamma_y = 1\times10^{-4}$) [23], respectively. Similarly, if Nernst effect was responsible for the measured second harmonic voltage, it should appear on the same order of magnitude in all three samples. The large difference shows that it should not be the dominant contribution. Therefore, the observed NPHE is likely to have a magnetic origin.

Ref. [31] reported a large hysteretic NPHE in magnetic TI $Cr_x(Bi,Sb)_{2-x}Te_3$/$(Bi,Sb)_2Te_3$ heterostructures and explains it as a result of asymmetric scattering of surface electrons by magnons. This magnon scattering mechanism cannot be applied to the TmIG/BS sample, because with OP magnetization, the magnons are polarized in the $z$-direction and cannot participate the scattering of surface electrons with IP-polarized spins. To see whether it is responsible for the hysteretic loop of the $R_{yx}^{2\omega}$ $vs$ $B$ curve in the YIG/BS sample, we parsed the NPHE into the linear-in-$B$ ($dR_{yx}^{2\omega}/dB$) and hysteretic ($\Delta R_{yx}^{2\omega}$) part, as displayed in Fig. 4(a). Fig. 4(b) shows that these two components have almost the same temperature dependence, suggesting that they share the same origin. Therefore, in the YIG/BS sample, it is likely that the IP exchange field experienced by the Dirac electrons from the IP magnetic moments of YIG plays the same role with the external magnetic field, which gives rise to the hysteretic loop around zero field. We note that although the easy axis of YIG is IP, the Dirac-electron-mediated exchange interaction [8,32] tilts the magnetic moments to the OP direction at the interface [8,32-34], which can also open an exchange gap for the TSS, just like that in TmIG/BS. As will be discussed in Section III.C, this gap opening may play a decisive role in generating the large NPHE in MI/TI heterostructures.

The other characteristics of the NPHE are shown in Fig. 5. First, the second-harmonic Hall resistance divided by field $dR_{ys}^{2\omega}/dB$ depends linear on the current density [Fig. 5(a) top inset], demonstrating its nonlinear nature. The deviation from linear relationship at higher current densities is due to Joule heating. Second, as shown in Fig. 5(a), when scanned under a $y$-direction field, $R_{yx}^{2\omega}$ becomes much smaller, consistent with the $\cos\phi_B$ dependence of NPHE. Third, we also measured the longitudinal second-harmonic resistance $R_{xx}^{2\omega}$ under a $y$-direction magnetic field scan and observed the longitudinal counterpart response, the so-called bilinear magnetoresistance (BMR) [35]. Lastly, as summarized in Fig. 5(b), the magnitude of the NPHE measure by $\gamma_y$ is enhanced at low carrier densities. This also points out the insignificant role of the hexagonal warping effect in the NPHE observed here, because hexagonal warping is negligible at low Fermi levels and is enhanced when carrier density increases.

## C. Discussion

The nonlinear charge current under IP electric and magnetic fields can be expressed as $j_a^{(2)} = \chi_{abcd} E_b E_c B_d$, where $\chi_{abcd}$ is the nonlinear conductivity tensor and $a, b, c, d = x$ or $y$. For an ideal linear Dirac dispersion, an IP magnetic field merely shifts the Dirac cone in $k$-space, leaving no effect on transport properties [22-24]. However, in real TIs, higher-order $k$ terms exist like the quadratic p-h asymmetry term and the cubic warping term. Here, for simplicity, we consider a linear TI dispersion with a parabolic $Dk^2$ term. As sketched in Fig. 6(a), when there is no IP magnetic field, under a driving electric field, there are equal number of electrons moving to the left and right with opposite spin polarizations. This generates a second-order spin current $j_s^{(2)}$ but there is no net charge current. When an IP magnetic field $B_x$ is applied [Fig. 6(b)], it not only shifts the Dirac cone, but also tilts it [23,24] in the $y$-direction due to the existence of the $k^2$ term. As a result, the transverse currents carried by the left- and right-moving electrons with opposite spin polarizations no longer cancel each other, and the nonlinear spin current is partially converted into a nonlinear charge current [22,35], giving rise to a nonlinear Hall conductivity $\chi_{yxxx}$. When the TI is further coupled to a magnetic material, the exchange interaction with OP moments opens a gap and introduces OP spin textures to the TSS [30]. Expanding around the shifted Dirac point, the Hamiltonian of TSS can be put in the form of $H = \hbar v_F (k_x \sigma_y - k_y \sigma_x) + \alpha B_x k_y + \Delta \sigma_z$ [23,24], where $\hbar$ is the reduced Planck constant, $v_F$ the Fermi velocity, and $\boldsymbol{\sigma}$ the Pauli matrices. The term with coefficient $\alpha B_x$ describes the magnetic-field-induced tilting in the $y$-direction, and $2\Delta$ is the OP exchange-interaction-induced gap. The IP magnetic field $B_x$ ~0.1 T used in this study is presumably smaller than the perpendicular magnetic anisotropy fields, so the size of the exchange gap is not affected by the small $B_x$. For similar tilted massive Dirac models, previous theoretical studies show that both intrinsic Berry-phase-related [36-39] and extrinsic skew-scattering or side-jump mechanisms [40-42] can contribute to the NPHE with different $\tau$ scaling. In our experiment, the one to two orders of magnitude increase of $\chi_{yxxx}$ ($\propto dR_{yx}^{2\omega}/IdB$) [Fig. 3] from 30 K to 4.5 K and the relatively small change of the linear conductivity in this regime [Fig. 2(a)] suggest that some other factor other than the relaxation time, governs the temperature dependence of the NPHE. As a result, we are not able to use $\tau$-scaling to narrow down the candidate NPHE mechanisms. More systematic future study is needed to clarify the dominant mechanism in the observed effect.

Our experimental results also suggest the importance of the OP spin textures formed in the TSS (due to MPE) for the NPHE, which was not considered in previous studies. As sketched in Fig. 6(c), with broken time-reversal symmetry, the spins of the electrons with $\pm k$ momenta are no longer orthogonal with each other, and backscattering between these states by nonmagnetic impurities is allowed. This is reflected as the suppression of WAL in the linear transport regime [Fig. 2(c)]. Ref. [23] shows that the tilting-induced PHE on the surface of TIs can be enhanced by scattering off nonmagnetic impurities. Here, we speculate that similar effect also occurs in the nonlinear regime. When backscattering is allowed due to OP spin texture formation, the nonlinear spin to charge current conversion may be increased, resulting in the large NPHE in MI/TI heterostructures. We note that in the above analysis, we only considered the TI surface states and did not consider the contribution from the bulk states. Because of the short-range nature of the MPE, the inversion symmetry of the bulk is presumably preserved. As a result, neither second-order spin nor charge current can be generated from the bulk [43].

## IV. CONCLUSION

In summary, we observed enhanced nonlinear planar Hall effect in $Bi_2Se_3$ films grown on magnetic TmIG and YIG substrates as compared to that on nonmagnetic GGG substrate. This NPHE is only observable below 30 K and scales inversely with carrier density. Compared with the previously reported NPHE in $Al_2O_3/Bi_2Se_3$ [22] arising from hexagonal warping or p-h asymmetry, the NPHE in MI/TI heterostructures is orders of magnitude larger, indicating a different origin. In YIG/$Bi_2Se_3$ we find the IP exchange field plays the same role as the external magnetic field, giving rise to an extra hysteresis loop in the $R_{yx}^{2\omega}$ $vs$ $B$ scans. Actually, a large hysteretic NPHE was previously observed in EuS/(Bi,Sb)$_2$Te$_3$ bilayers in Ref. [44], and a mechanism based on tilting of the p-h asymmetric Dirac cone by the IP exchange field was speculated. Our control experiments suggest the necessary role of the OP exchange-interaction-induced gap opening or modified spin textures in generating the large NPHE. Further experimental and theoretical work is needed to reveal the underlying mechanism and establish the NPHE as a convenient and sensitive transport probe of the MPE in MI/TI heterostructures.

### ACKNOWLEDGMENTS

This work was supported by the U.S. DOE, Office of Basic Energy Sciences under Contract No. DE-SC0016380 and by NSF DMR Grant No. 1904076. Y.-X. H and S. A. Yang are supported by Singapore NRF CRP22-2019-0061. C. X. is supported by the UGC/ RGC of Hong Kong SAR (AoE/P-701/20). The authors acknowledge the use of the Materials Growth Facility (MGF) at the University of Delaware, which is partially supported by the National Science Foundation Major Research Instrumentation under Grant No. 1828141 and UD-CHARM, a National Science Foundation MRSEC, under Award No. DMR-2011824.

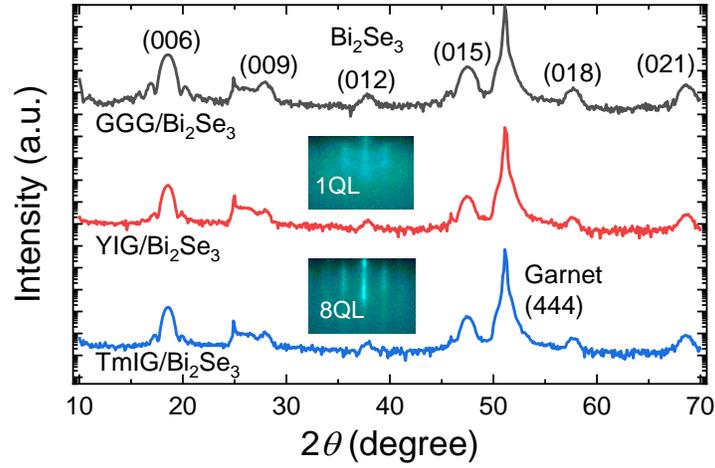

FIG. 1 XRD and RHEED results of Bi$_2$Se$_3$(8 QL) films grown on TmIG, YIG and GGG substrates.

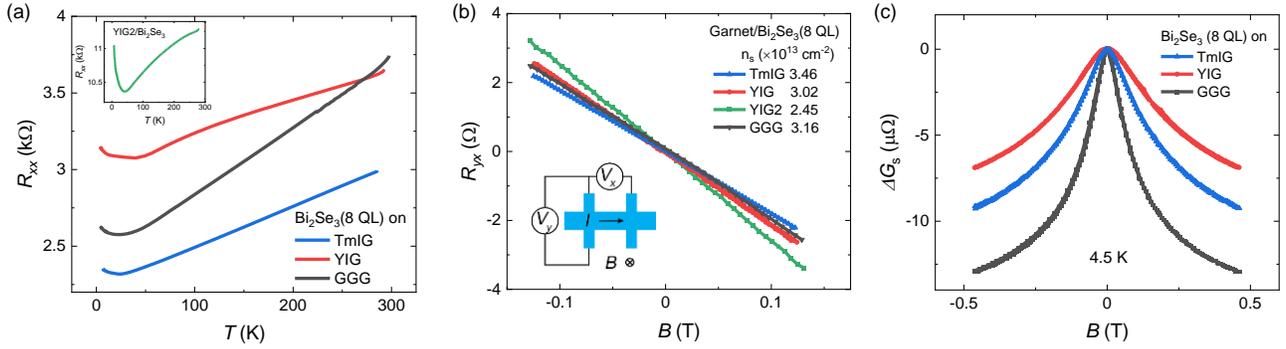

FIG. 2 (a) Temperature dependence of the longitudinal resistance $R_{xx}$ of four Garnet/Bi$_2$Se$_3$(8 QL) samples. YIG and YIG2 are two films with thicknesses 2.5 $\mu$m and 100 nm respectively. The result of YIG2/BS is plotted in the inset due to its large resistance. (b) Hall resistance as a function of OP magnetic field and extracted sheet carrier densities. Inset: Schematic of the measurement setup. (c) Change of sheet conductance under an OP magnetic field for three Garnet/Bi$_2$Se$_3$(8 QL) devices at 4.5 K.

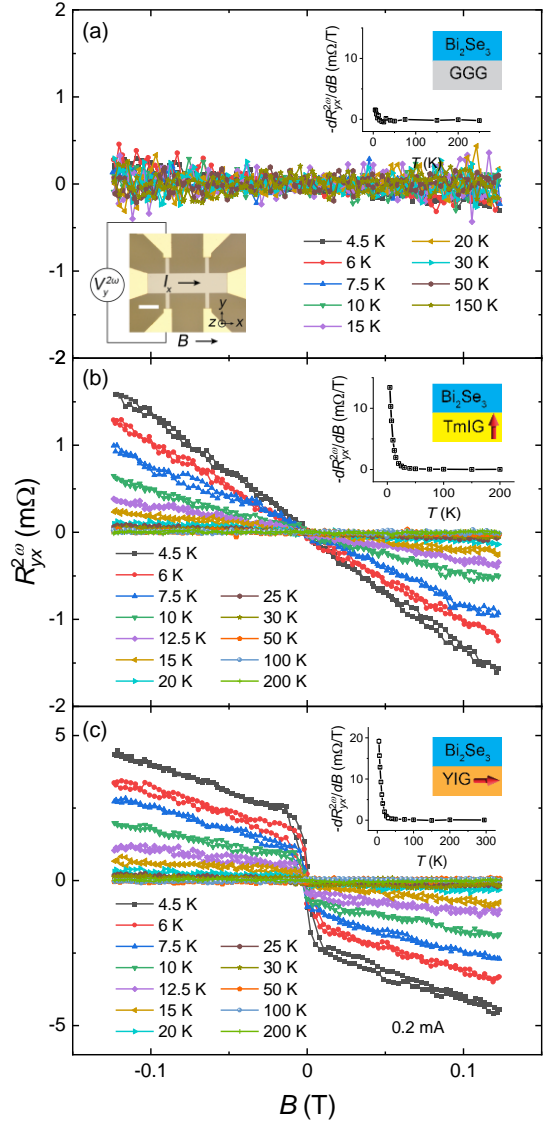

FIG. 3 The nonlinear Hall resistance $R_{yx}^{2\omega} = V_y^{2\omega}/I$ as a function of $x$-direction magnetic field for three $Bi_2Se_3$ samples grown on (a) GGG, (b) TmIG, and (c) YIG substrates at various temperatures. The bottom inset in (a) is an optical image of a device with illustrated measurement setup. Scale bar, 100 $\mu$m. The top insets in (a)-(c) display the temperature dependence of the slope $-dR_{yx}^{2\omega}/dB$.

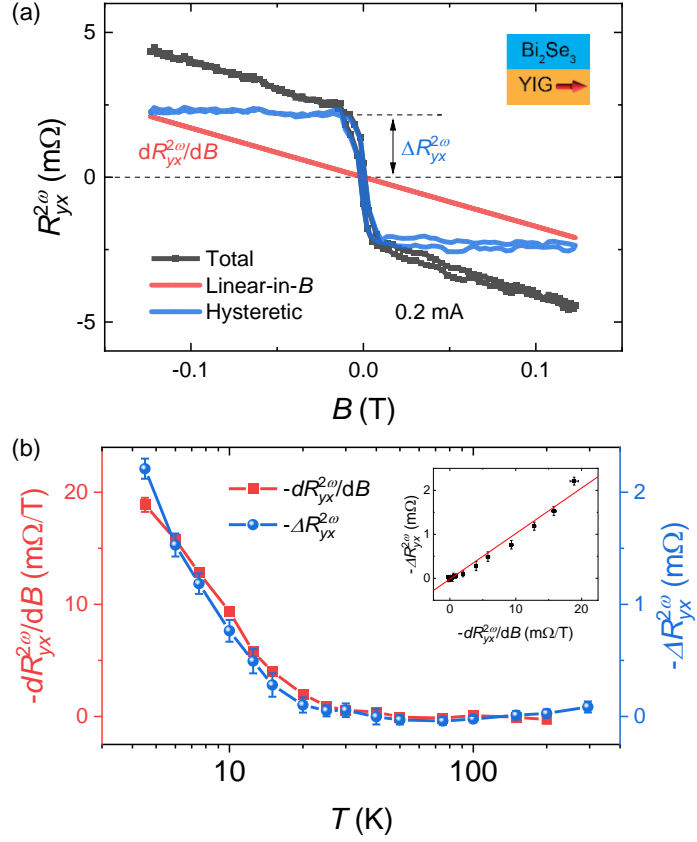

FIG. 4 (a) The $R_{yx}^{2\omega}$ vs $B$ curve (black) of the YIG/BS sample consists of a linear-in-$B$ component (red) with slope $dR_{yx}^{2\omega}/dB$ and a hysteretic component (blue) with magnitude $\Delta R_{yx}^{2\omega}$. (b) Temperature dependence of $-dR_{yx}^{2\omega}/dB$ and $-\Delta R_{yx}^{2\omega}$. Inset shows the linear relationship between them.

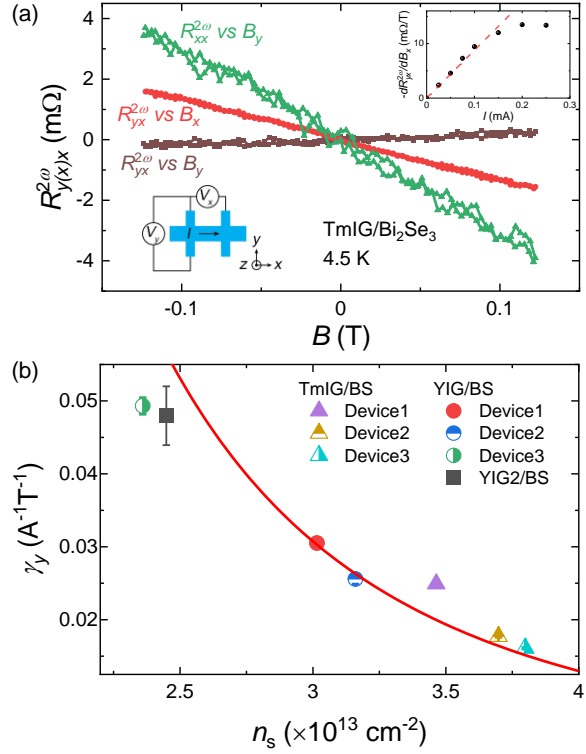

FIG. 5 (a) Longitudinal and transverse second-harmonic resistance as a function of IP $x$- or $y$-direction magnetic field for the TmIG/Bi$_2$Se$_3$ sample. Top inset: Current density dependence of the nonlinear Hall resistance $-dR_{yx}^{2\omega}/dB$. Bottom inset: Schematic of the measurement setup. (b) Carrier density dependence of the coefficient $\gamma_y = \frac{R_{yx}^{2\omega}}{R_{xx}IB}$ in several devices. The solid line fit to $n_s^{-3}$ is guide for the eye.

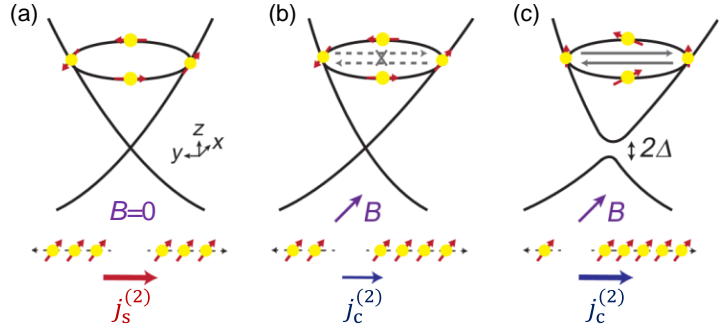

FIG. 6 Illustration of the hypothetical backscattering enhanced NPHE mechanism. (a) For a TI surface dispersion containing a parabolic term, when there is no IP magnetic field, under a driving electric field, the number of electrons traveling to the right carrying up spins is equal to that of the electrons traveling to the left with down spins. This generates a second-order spin current $j_s^{(2)}$ but there is no charge current. (b) An IP magnetic field tilts the upright Dirac cone and causes an imbalance between the left- and right-moving electrons, resulting in a second-order charge current, i.e., the NPHE. (c) When the TI is further coupled with a magnetic material, the exchange interaction opens an exchange gap $2\Delta$ and introduces OP spin polarization components to the surface electrons. The restriction on backscattering is lifted, which may enhance the nonlinear spin to charge conversion efficiency, giving rise to the large NPHE in MI/TI heterostructures.

13